\documentclass[final,5p,times,twocolumn]{elsarticle}

\usepackage{amssymb}
\usepackage{lipsum}
\journal{Physics Letters B}
\pdfoutput=1
\usepackage{tabularx}
\usepackage[T1]{fontenc}
\usepackage[english]{babel}
\usepackage{amsmath}
\usepackage{graphicx}
\usepackage{dcolumn}
\usepackage{pbox}
\usepackage{amssymb}
\usepackage{epsfig}
\usepackage{slashed}
\usepackage{amssymb}
\usepackage{ mathrsfs }
\usepackage{float}
\usepackage{natbib}
\usepackage[usenames,dvipsnames]{xcolor}
\usepackage[font=small]{caption}
\usepackage[font=small]{subcaption}
\usepackage{url}
\def\simlt{\stackrel{<}{{}_\sim}}
\def\simgt{\stackrel{>}{{}_\sim}}
\usepackage{multirow}
\usepackage{lineno}
\usepackage[figuresright]{rotating}

\definecolor{MyLightBlue}{rgb}{0.22,0.51,0.9}
\definecolor{BrickRed}{rgb}{0.8, 0.25, 0.33}
\RequirePackage{hyperref}
\hypersetup{colorlinks, citecolor=blue,linkcolor=BrickRed, urlcolor=MyLightBlue}
\biboptions{comma,square,compress,numbers}
\begin{document}
\title{Doubly-charged scalars of the Minimal Left-Right Symmetric Model at Muon Colliders}
\author[first]{Mohamed Belfkir}
\ead{mohamed.belfkir@cern.ch}
\address[first]{Department of Physics, UAE University, P.O. Box 17551, Al-Ain, United Arab Emirates}
\author[second,third,fourth]{Talal Ahmed Chowdhury}
\ead{talal@du.ac.bd}
\address[second]{Department of Physics, University of Dhaka, Dhaka 1000, Bangladesh}
\address[third]{Department of Physics and Astronomy, University of Kansas, Lawrence, Kansas 66045, USA}
\address[fourth]{The Abdus Salam International Centre for Theoretical Physics, Strada Costiera 11, I-34014, Trieste, Italy}
\author[first,fourth]{Salah Nasri}
\ead{snasri@uaeu.ac.ae}

\begin{abstract}
    We investigate the prospects of probing the doubly-charged scalars of the minimal Left-Right Symmetric model (MLRSM) at a muon collider. We assess its capability by studying the production of doubly-charged scalars and their subsequent decay into four charged lepton final states containing the same-charge lepton pairs. We find that the channels with same-charge electron and muon pairs, i.e., ($e^{\pm}e^{\pm}\mu^{\mp}\mu^{\mp}$ and its charge conjugated pairs), have the largest sensitivity due to the lowest Standard Model background. Besides, we show that the possibility of using fully polarized initial muon beams in the muon collider can enhance the detection sensitivity of doubly-charged scalars of the MLRSM. Furthermore, we show that one can put exclusion limits on the magnitudes of triplet Yukawa couplings that are directly related to the neutrino sector of the MLRSM for the mass range $1.1-5$ TeV of the doubly-charged scalars.
\end{abstract}

\maketitle

\section{Introduction}\label{intro}
The minimal left-right symmetric model (MLRSM) \cite{Mohapatra:1974gc, Mohapatra:1974hk, Senjanovic:1975rk, Senjanovic:1978ev, Mohapatra:1979ia, Mohapatra:1980yp} is a compelling extension of the Standard Model (SM) that addresses the fundamental question of neutrino mass. Neutrino mass is a significant departure from the SM, and uncovering its underlying mechanisms is crucial to unveil new physics beyond the SM. The MLRSM incorporates the seesaw mechanism \cite{Minkowski:1977sc, Mohapatra:1979ia, Gell-Mann:1979vob, Glashow:1979nm, Yanagida:1979as}, which has emerged as the prevailing explanation for the smallness of neutrino mass. The MLRSM, however, goes beyond the seesaw mechanism's ad hoc nature and offers a more comprehensive framework. It encompasses the left-right symmetry \cite{Pati:1974yy}, which attributes the left-handed nature of weak interactions to the spontaneous breakdown of parity. The model postulates the existence of right-handed (RH) neutrinos ($\nu_R$), thereby providing a natural explanation for non-vanishing neutrino masses. In the minimal version of the MLRSM, additional Higgs scalars in the form of left-handed and right-handed triplets play a vital role in spontaneous symmetry breaking\footnote{Additionally, it was shown in \cite{Chang:1983fu} that the right-handed triplets are sufficient to break the symmetry and the left-handed triplets can be kept at higher scale.}. Significantly, the mass of the RH-charged gauge boson ($W_R$) is directly related to the mass of the RH neutrinos, $M_N$, reinforcing the connection between the smallness of the neutrino mass and the parity violation in weak interactions.

Besides, the MLRSM is emerging as a self-contained and predictive theory of neutrino mass like the Higgs origin of the charged fermion masses in the SM \cite{Nemevsek:2012iq, Senjanovic:2016vxw, Senjanovic:2018xtu, Senjanovic:2019moe, Senjanovic:2023czt}. Moreover, the MLRSM not only predicts the lepton number violation both at low energies through the neutrinoless double beta decay \cite{Racah:1937qq, Furry:1939qr} and at high energies through the Keung-Senjanovic process, i.e., the production of same-sign charged lepton pairs in hadron colliders \cite{Keung:1983uu} but also connects these low-energy and high-energy processes \cite{Tello:2010am, Nemevsek:2011aa, Chakrabortty:2012mh}. Moreover, despite having near maximal parity violation in the low energy weak interactions, the left and right quark mixing matrices are shown to be close to each other \cite{Zhang:2007fn, Senjanovic:2015yea}. The MLRSM has been studied extensively in the Large Hadron Collider (LHC), and in \cite{Nemevsek:2018bbt, CMS:2021dzb, ATLAS:2023cjo}, it was shown that for the mass of the RH neutrinos, $M_{N}\simlt 1$ TeV, the observed lower limit on the mass of the RH gauge boson, $W_{R}$ is $M_{W_{R}}>6.4$ TeV. Also, in \cite{ATLAS:2022pbd}, it was shown that the observed lower limit on the mass of the doubly-charged scalars in the MLRSM is 1.08 TeV.

Recently, the muon colliders are considered to be future lepton colliders as these circular colliders have the potential to achieve center-of-energy in the multi-TeV range with high luminosity \cite{Gunion:1998bc, Palmer:2014nza}. In addition, the muon collider is a promising lepton collider for carrying out precision studies on the Higgs, gauge boson and Yukawa sectors of the SM as well as probing different BSM scenarios (see for example, \cite{AlAli:2021let, Aime:2022flm}). Therefore, it will be interesting to study the sensitivity of such a future muon collider in probing the BSM scalars and gauge bosons predicted by the MLRSM. In this work, we have studied the sensitivity of a muon collider to probe the LH and RH doubly-charged scalars, $\Delta_{L}^{++}$ and $\Delta^{++}_{R}$, respectively, which are uniquely predicted by the MLRSM. Moreover, we have shown that the possibility of using the polarized incoming muon beams, which can be a distinctive feature of the future muon collider, will distinguish between LH and RH doubly charged scalars of the MLRSM.

The article is organized as follows. In section~\ref{mlrsmsec} we present the model. In section~\ref{tripletmuoncollider} we discuss the relevant parameter space of the model, identify the signals of interest and describe the event reconstruction and selection at the muon collider. Section~\ref{resultsec} includes the results of our analysis. Finally, we conclude in section~\ref{conclusion}.

\section{Doubly-charged scalars of the MLRSM}\label{mlrsmsec}
The gauge group of the MLRSM is $SU(3)_{c}\times SU(2)_{L}\times SU(2)_{R}\times U(1)_{B-L}$ with a symmetry between the left and right sectors given by the generalized parity ($\mathcal{P}$) or charge conjugation ($\mathcal{C}$) symmetry. Quarks and leptons are represented under the MLRSM gauge group as,
\begin{equation}
    Q_{L,R}=\begin{pmatrix}
        u\\
        d
    \end{pmatrix}_{L,R},\,\,\,
    \ell_{L,R}=\begin{pmatrix}
        \nu\\
        e
    \end{pmatrix}_{L,R}.
\label{lrsm-1}
\end{equation}
Moreover, the electromagnetic charge operator is defined as, $\hat{Q}_{\mathrm{EM}}=T^{3}_{L}+T^{3}_{R}+\frac{\hat{Q}_{BL}}{2}$, where $T^{3}_{L,R}$ are the diagonal generators of $SU(2)_{L,R}$, respectively, and $\hat{Q}_{BL}$ gives the charge under $U(1)_{B-L}$. For quark and lepton doublets, it is simply $\hat{Q}_{BL}=1/3$ and $-1$, respectively.
\begin{figure}[t!]
\vspace{-1cm}
\centerline{
\includegraphics[width=0.25\textwidth]{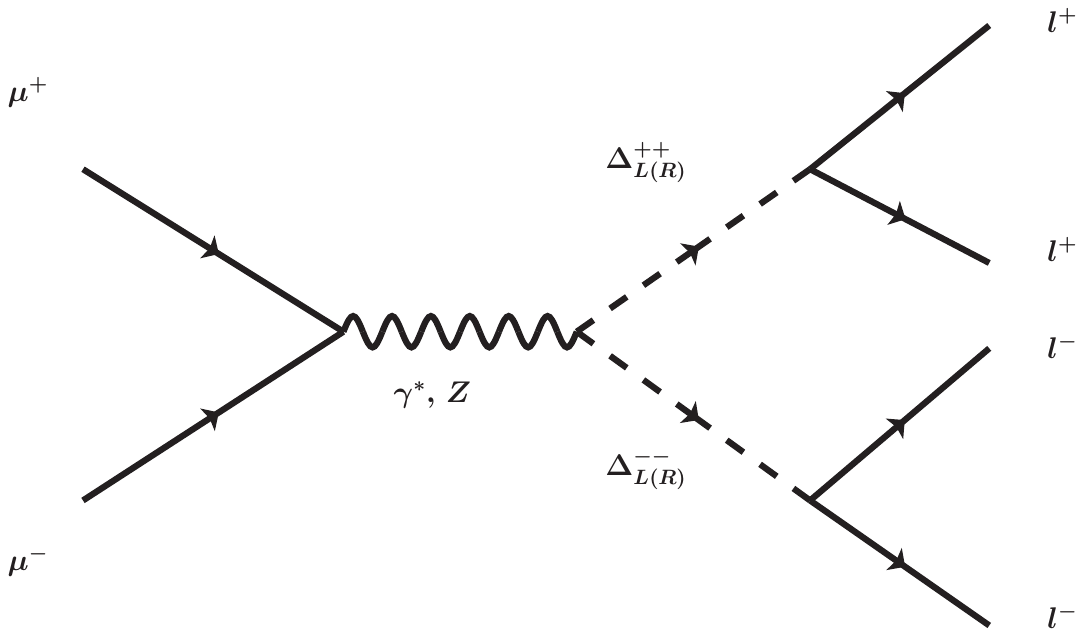}\hspace{1mm}
\includegraphics[width=0.25\textwidth]{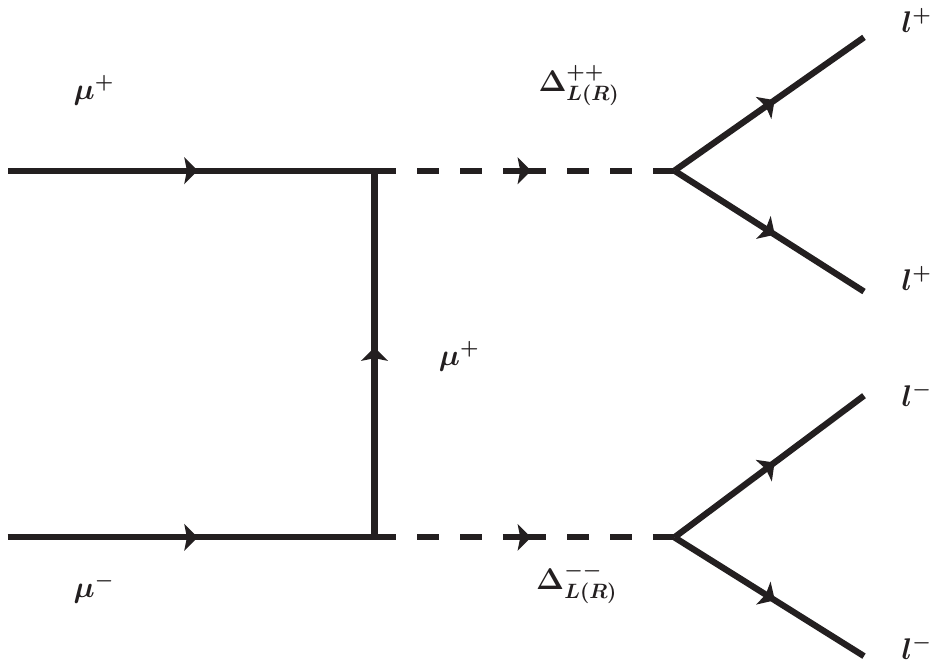}}\vspace{-3.5cm}
\caption{The Feynman diagrams corresponding to the dominant processes which contribute to $\mu^{+}\,\mu^{-}\rightarrow \Delta^{++}_{L(R)}\,\Delta^{--}_{L(R)}\rightarrow \ell^{+}\ell^{+}\ell^{-}\ell^{-}$.}
\label{feynmandig}
\end{figure}
The Higgs sector of the MLRSM \cite{Mohapatra:1979ia, Mohapatra:1980yp, Deshpande:1990ip, Duka:1999uc, Barenboim:2001vu, Kiers:2005gh, Tello:2012qda, Dev:2016dja, Maiezza:2016bzp, Maiezza:2016ybz, BhupalDev:2018xya} contains a bidoublet $\Phi$ and $SU(2)_{L,R}$ scalar triplets, $\Delta_{L,R}$ under the gauge group, as follows,
\begin{equation}
\Phi \in (1,2,2,0),\,\,\,\mathbf{\Delta}_{L}\in (1,3,1,2)\,\,\, \mathrm{and}\,\,\, \mathbf{\Delta}_{R}\in (1,1,3,2),
\label{lrsm-2}
\end{equation}
where,
\begin{equation}
    \Phi=\begin{pmatrix}
        \phi^{0}_{1} & \phi^{+}_{2}\\
        \phi^{-}_{1} & \phi^{0}_{2}
    \end{pmatrix},\,\,\,
    \mathbf{\Delta}_{L,R}=\begin{pmatrix}
        \Delta^{+}/\sqrt{2} & \Delta^{++}\\
        \Delta^{0} & -\Delta^{+}/\sqrt{2}
    \end{pmatrix}_{L,R}.
    \label{lrsm-3}
\end{equation}
The Yukawa sector is given by,
\begin{align}
\mathcal{L}_{Y}&= \overline{Q_{L}}\left(Y_{q}\,\Phi+\tilde{Y}_{q}\,\tilde{\Phi}\right)Q_{R}+\overline{\ell_{L}}\left(Y_{\ell}\,\Phi+\tilde{Y}_{\ell}\,\tilde{\Phi}\right)\ell_{R}\nonumber\\
&+\frac{1}{2}\left(\overline{(\ell_{L})^{c}}\,\epsilon\,Y_{L}\mathbf{\Delta}_{L}\ell_{L}+\overline{(\ell_{R})^{c}}\,\epsilon\,Y_{R}\mathbf{\Delta}_{R}\ell_{R}\right)+\mathrm{h.c},
\label{lrsm-4}
\end{align}
where, $\epsilon=i\sigma_{2}$, $\tilde{\Phi}=\epsilon\Phi\epsilon$ and $(f_{L(R)})^{c}=C\gamma^{0}f^{*}_{L(R)}$ is the usual charge-conjugate spinor.
Now, under the discrete left-right symmetry, the fields transform as follows:
\begin{align}
    \mathcal{P}:&\,\,\,f_{L}\longleftrightarrow f_{R},\,\,\,\Phi\longleftrightarrow \Phi^{\dagger},\,\,\,\mathbf{\Delta}_{L}\longleftrightarrow \mathbf{\Delta}_{R},\nonumber\\
    \mathcal{C}:&\,\,\,f_{L}\longleftrightarrow (f_{R})^{c},\,\,\,\Phi\longleftrightarrow \Phi^{T},\,\,\,\mathbf{\Delta}_{L}\longleftrightarrow \mathbf{\Delta}^{*}_{R},\label{lrsm-5}
\end{align}
and the invariance of the Lagrangian under the left-right symmetry imposes the following conditions on the Yukawa couplings as follows:
\begin{align}
\mathcal{P}:&\,\,\,Y_{q}=Y_{q}^{\dagger},\,\,Y_{l}=Y_{l}^{\dagger},\,\,Y_{L}=Y_{R},\\
\mathcal{C}:&\,\,\,Y_{q}=Y_{q}^{T},\,\,Y_{l}=Y_{l}^{T},\,\,Y_{L}=Y^{*}_{R}.
\end{align}
In this work, we consider the generalized parity $\mathcal{P}$ as the discrete left-right symmetry, so the Yukawa couplings relevant for the scalar triplets become $Y_{L}=Y_{R}=Y$. Besides, the gauge couplings associated with $SU(2)_{L,R}$ are $g_{L}=g_{R}=g$. We also listed the parity symmetric Higgs potential in the appendix.

The symmetry breaking in the MLRSM takes place in two steps. First, at the high scale with the breaking of $SU(2)_{R}\times U(1)_{B-L}\rightarrow U(1)_{Y}$ through the vacuum expectation values of scalar triplets,
\begin{equation}
\langle \Delta^{0}_{R}\rangle =v_{R}/\sqrt{2},\,\,\,\langle \Delta^{0}_{L}\rangle = v_{L}/\sqrt{2},
\label{lrsm-6}
\end{equation}
where we consider $v_{L}\sim 0$, which is set by the seesaw picture of the MLRSM, as it directly contributes to the Majorana mass of the $\nu_{L}$. Afterwards, at the lower scale, the SM symmetry is broken by the vev of $\Phi$, which is expressed as, 
\begin{equation}
    \langle\Phi\rangle=v/\sqrt{2}\,\,\mathrm{diag}(\cos\beta, -e^{i a}\sin\beta) 
\end{equation}
where, $\tan\beta=v_{2}/v_{1}$, the ratio between vevs of $\langle\phi_{1,2}^{0}\rangle=v_{1,2}$, $v=\sqrt{v_{1}^2+v_{2}^2}$ being the SM vev of 246 GeV, and the presence of the phase $a$ leads to the spontaneous breaking of CP symmetry by this vev.

Now, as $v_{R}\gg v$, neglecting the $O(v/v_{R})$ terms, the masses of the charged gauge boson, $W_{R}$ and neutral gauge boson, $Z_{R}$ are given by,
\begin{equation}
m_{W_{R}}^{2}\simeq \frac{1}{2} g^2 v_{R}^2,\,\,\, m_{Z_{R}}^{2}\simeq \left(g^2+\frac{1}{2}g^{2}_{B-L}\right)v_{R}^2,
\label{lrsm-7}
\end{equation}
where, the gauge coupling $g_{B-L}$ can be determined using the relation,
$1/e^2=2/g^2+1/g^{2}_{B-L}$. Moreover, the RH neutrino mass matrix is $M_{\nu_{R}}=Y\,v_{R}/\sqrt{2}$, which is proportional to the $v_{R}$ and connects the neutrino mass to the symmetry breaking scale of the MLRSM.

We will focus on the scalar triplets, specifically the doubly-charged scalars $\Delta^{++}_{L,R}$ of the MLRSM. As $v_{L}\sim 0$, the decay $\Delta^{\pm\pm}_{L}\rightarrow W^{\pm}\,W^{\pm}$ to the same-sign SM charged gauge bosons $W$ is negligible. Now, the presence of flavor changing neutral currents mediated by the scalars, $H^{+},\,H^{0}\,A^{0}$ coming from the bidoublet set their masses to be $\simgt 20$ TeV, which in turn puts the $W_{R}$ mass (or the vev $v_{R}$) also in the multi-TeV range~\cite{Beall:1981ze, Mohapatra:1983ae, Ecker:1983uh, Zhang:2007da, Maiezza:2010ic, Guadagnoli:2010sd, Bertolini:2014sua, Bertolini:2019out}. But the masses of the $\Delta^{++}_{L}$, $\Delta^{+}_{L}$, $\Delta^{0}_{L}$, $\Delta^{++}_{R}$ and $\mathrm{Re}\Delta^{0}_{R}$ (as seen in appendix), can be set to a smaller scale than the $m_{W_{R}}$ in a region of the parameter space of the MLRSM. Therefore, we focus on this parameter space where, apart from the above-mentioned scalars associated with scalar triplets, the BSM particles predicted by the MLRSM are in the $O(\simgt 20)$ TeV range. Hence, $\Delta^{\pm\pm}_{R}\rightarrow W^{\pm}_{R}\,W^{\pm}_{R}$, to the same-sign $W_{R}$ becomes kinematically forbidden in our case. Consequently, the dominant decay modes of the doubly-charged scalars turn out to be the same-sign charged lepton pairs.

\section{Doubly-charged scalars at the muon collider}\label{tripletmuoncollider}
\subsection{The parameter space}
The parameter space of the MLRSM associated with its scalar sector is large. For this reason, one needs to focus on particular benchmarks to study the detection possibility in a collider. In this regard, as a most simplest option, we consider the parameter space where the component fields of the $\mathbf{\Delta}_{L}$ has near mass degeneracy: $m_{\Delta^{++}_{L}}\simeq m_{\Delta^{+}_{L}}\simeq m_{\Delta^{0}_{L}}$. Additionally, we consider the degeneracy of LH and RH doubly charged scalars, $m_{\Delta_{L}^{++}}\simeq m_{\Delta^{++}_{R}}$, aligning our analysis with ATLAS searches reported in \cite{ATLAS:2022pbd}. In this chosen parameter region, as the mass difference between $\Delta^{++}_{L}$ and $\Delta^{+}_{L}$ are negligible, the cascade decay, $\Delta^{++}_{L}\rightarrow \Delta^{+}_{L}W^{+}$ is suppressed. Therefore, we only focus on the dominant decay modes of doubly charged scalars are into the same-charge lepton pairs,
\begin{equation}
    \Delta^{\pm\pm}_{L(R)}\rightarrow \ell^{\pm}_{L(R)}\ell'^{\pm}_{L(R)}\,\,\,\mathrm{with}\,\,\,\Gamma^{\Delta^{++}}_{\ell\ell'}=\kappa \frac{|Y_{\ell\ell'}|^{2}}{16\pi}m_{\Delta^{++}},
    \label{lrsm-8}
\end{equation}
where $\kappa=2$ for $\ell=\ell'$ and $\kappa=1$ for $\ell\neq\ell'$, respectively. Unlike the case of hadron colliders, in the muon collider, the production of the doubly-charged scalars proceeds through the Drell-Yan-like processes (s-channel) with $\gamma^*$ and $Z$, $Z_{R}$ and heavy Higgses, and additional t-channel processes with the exchange of charged leptons. But, as $Z_{R}$ and heavy Higgses have masses in $O(20\,\mathrm{TeV})$, their contributions will be sub-dominant compared to $\gamma^*$ and $Z$ for our considered center-of-mass energy $\sqrt{s}\leq 10$ TeV. Therefore, our analysis focuses on s-channel processes with $\gamma^*$ and $Z$, as shown in Fig.~\ref{feynmandig} (left), and t-channel processes with the exchange of charged leptons, as shown in Fig.~\ref{feynmandig} (right). In addition, the amplitude associated with Fig.~\ref{feynmandig} (left) scales with the couplings $\sim O(g^2 Y^{2})$ whereas the t-channel process given in Fig.~\ref{feynmandig} (right) has $O(Y^{4})$ dependence, where $Y$ denotes the Yukawa couplings between the doubly-charged scalars and charged leptons. For this reason, if $Y_{\ell\ell}\simgt g$, the t-channel dominates the pair productions of doubly-charged scalars and their subsequent decays into same-charge lepton pairs. But the scale of these Yukawa couplings, particularly the off-diagonal ones, are severely constrained by the limits on the charged lepton flavor violating (LFV) processes: $\mathrm{Br}(\mu\rightarrow 3e)< 10^{-12}$ \cite{SINDRUM:1987nra}, $\mathrm{Br}(\tau\rightarrow 3 e)<2.7\times 10^{-8}$ and $\mathrm{Br}(\tau\rightarrow 3 \mu)< 2.1\times 10^{-8}$ \cite{Hayasaka:2010np}, as the doubly-charged scalars mediate them at the tree-level (see for example~\cite{Cirigliano:2004mv}). For this reason, we consider the Yukawa coupling matrix $Y$ to be diagonal for simplicity, i.e., $Y_{ee}=Y_{\mu\mu}=Y_{\tau\tau}$ and $Y_{\ell\ell'}=0$. Thus branching ratios to each of the possible leptonic final states of the doubly-charged scalars are set equal to 1/3. Moreover, as the muon collider is expected to have center-of-mass energy $\sqrt{s}=10$ TeV with an integrated luminosity of $\mathcal{L}=10\,\mathrm{ab}^{-1}$, we focus on the mass range of the doubly-charged scalars in $m_{\Delta^{++}}=1100 - 5000\,\mathrm{GeV}$.

\begin{figure}[t!]
    \centerline{
    \subfloat[(a)][]{\includegraphics[width=0.27\textwidth]{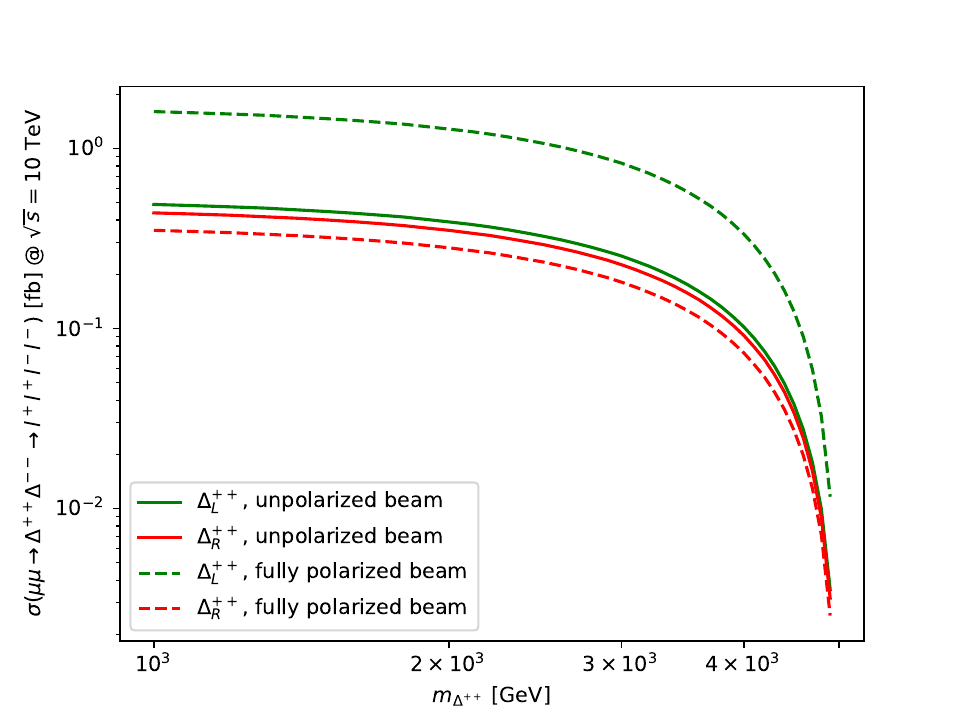}}
    \subfloat[(b)][]{\includegraphics[width=0.27\textwidth]{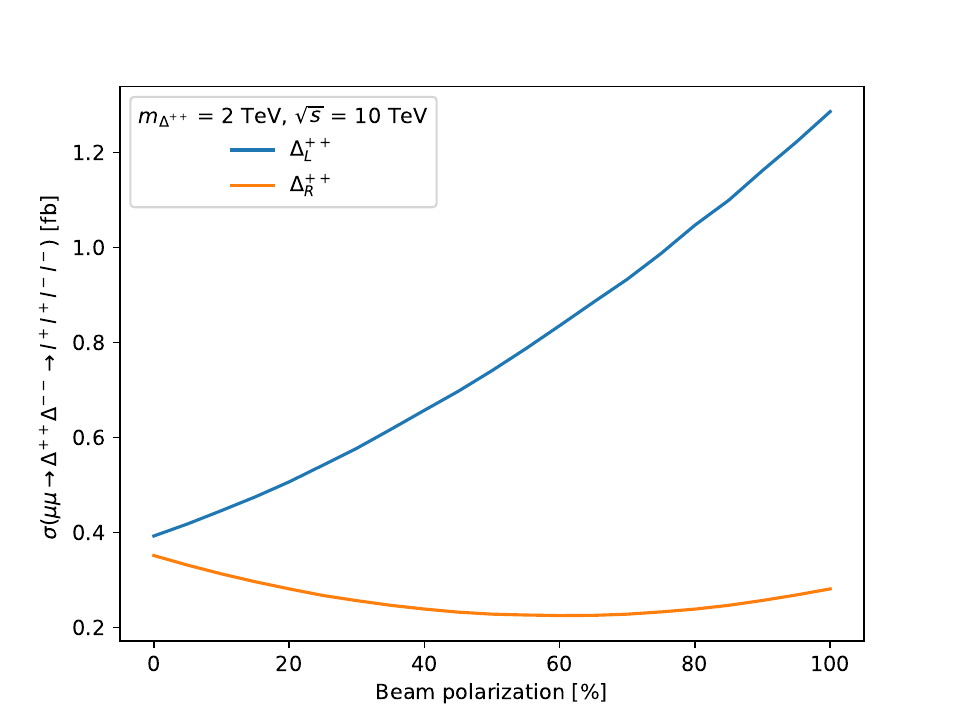}}}
    \caption{Production cross-section (a) as a function of scalar mass for unpolarized beam (polarization = 0\%) and fully polarized beam (polarization = 100\%), (b) as a function of the initial beam polarization fraction for a scalar mass of 2 TeV. All the Yukawa couplings are set to be 0.01. }
    \label{fig:cross_vs_pol}
\end{figure}

Fig.~\ref{fig:cross_vs_pol} (a) shows the total cross-section of the doubly-charged scalar pair productions decaying to four leptons $\sigma(\mu\mu\rightarrow\Delta^{++}_{L,R}\Delta^{--}_{L,R}\rightarrow \ell^{+}\ell^{+}\ell^{-}\ell^{-})$ ($\ell= e$ or $\mu$) as a function of the scalar mass for $\sqrt{s}$ = 10 TeV. The cross-section drops significantly, as expected, at around $\sqrt{s}/2$. The $\Delta^{++}_{L}$ shows a slightly higher production rate compared to the $\Delta^{++}_{R}$, which can be enhanced with the polarized initial muon beam. Actually, $\Delta^{++}_{L}$ and $\Delta^{++}_{R}$ couple to the  $Z$ bosons with $\frac{g}{\cos\theta_{w}}(1-2\sin^{2}\theta_{w})$ and $-2 g \sin\theta_{w}\tan\theta_{w}$, respectively. On the other hand, if the initial muon beams are polarized, i.e., LH and RH muons, the $Z$ boson couplings are $\frac{g}{\cos\theta_{w}}(-\frac{1}{2}+\sin^{2}\theta_{w})$ and $g\sin\theta_{w}\tan\theta_{w}$, respectively. Here $\theta_{w}$ is the Weinberg angle. So using the polarized initial muon beam would lead to distinguishing the Drell-Yan process involving $Z$ boson as the t-channels are highly suppressed for the polarized cases. Fig.~\ref{fig:cross_vs_pol} (b) shows the production cross-section in the four-lepton final state as a function of the initial muon beam polarization. The beam polarization refers to the degree of alignment of the muon spins within the beam line. We can see from Fig.~\ref{fig:cross_vs_pol} (b) that for $m_{\Delta^{++}}$ = 2 TeV, a fully polarized beam leads to about a factor 4 enhancement in the $\Delta^{++}_{L}$ production cross-section compared to $\Delta^{++}_{R}$, thus making $\Delta^{++}_{L}$ and $\Delta^{++}_{R}$ more distinguishable in the muon collider.

\subsection{Monte Carlo event generation}
\label{mc}

The signal of interest consists of final states with four leptons. The doubly-charged scalars decay into the same-charge lepton pairs, leading to three different signal categories \footnote{As the $\tau$ decay modes are more involved, $\tau$ in the final states are not considered in this analysis.}. The same flavor - same flavor ($SF-SF$) where doubly-charged scalars decay into the same flavor lepton pairs. This category includes final states with four muons  ($\mu^{+}\mu^{+}\mu^{-}\mu^{-}$), four electrons ($e^{+}e^{+}e^{-}e^{-}$), and two electrons and two muons ($e^{\pm}e^{\pm}\mu^{\mp}\mu^{\mp}$ and its charge conjugated states). We denote these four lepton combinations as $4\mu$, $4e$, and $2e 2\mu$, respectively, in the subsequent text. The second category is the same flavor - different flavor ($SF-DF$),  where one of the doubly-charged scalars decays into the same flavor lepton pair, while the other decays into different flavor pair. This category includes final states with three electrons and one muon ($eee\mu$) and three muons and one electron ($e\mu\mu\mu$). The third category consists of different flavor - different flavor ($DF-DF$) decay, where doubly-charged scalars decay into opposite flavor pairs. This category includes final states with two electrons and two muons of the form ($e\mu e\mu$). Only the $SF-SF$ final state is analyzed, the $SF-DF$ and $DF-DF$ scenarios are not considered in this work due to the mixing between electrons and muons being set to zero. 

For each considered $SF-SF$ leptonic final state, 50,000 events were generated at a center-of-mass energy of 10 TeV, corresponding to an integrated luminosity of 10 $ab^{-1}$, as defined by the Snowmass muon collider forum \cite{Aime:2022flm}. We also consider the use of the muon beam polarization to enhance the search sensitivity. Two different beam polarization configurations are adopted. The No polarization, where the muon beams are considered unpolarized or the fully polarized beam, where the positively charged muon beam is polarized right-handed and the negatively charged muon beam is considered with left-handed polarization.

The simplified MLRSM lagrangian (Equation \ref{lrsm-4}) is implemented using the \texttt{Feynrules} package \cite{Alloul:2013bka} to generate the UFO files. Afterwards, the event generation is performed at the leading-order (LO) matrix elements with \texttt{MadGraph5\_aMC\@NLO 3.3.0} \cite{Alwall:2014hca}, which is interfaced to \texttt{Pythia 8.186} \cite{Bierlich:2022pfr} for decay chain modelling, parton showering, hadronization, and the description of the underlying event. The dominant Standard Model (SM) background for most channels arises from misidentifying lepton flavors in the final states. The misidentification rate for electrons and muons in the muon colliders is estimated to be less than 0.5\%, making the final states nearly background-free \cite{Yu:2017mpx}. However, the SM background that mainly includes $ZZ^{*}$, $Z\gamma$, and $\gamma\gamma$ processes \footnote{in addition to four lepton decay}, was generated with a total of 100,000 events for each channel ($4\mu$, $4e$, and $2e2\mu$), following the same signal generation scheme.

The \texttt{Delphes} \cite{deFavereau:2013fsa} fast simulation is used to emulate the detector reconstructions and performances. Moreover, the muon collider detector card is included in its latest release \cite{delphs_muon_card}, which is a hybrid of CLIC \cite{leogrande2019delphes} and FCC-$hh$ cards \cite{Selvaggi:2717698}. The card assumes a muon (electron) reconstruction efficiency of nearly 100\% (90\%) for $|\eta| < 1.5$, and 98\% (75\%) for $1.5 < |\eta| < 2.5$. The muon $p_T$ resolution is approximately 1\%. Both signal and background are passed through the same reconstruction process. 

\subsubsection{Events reconstruction and selection}
Leptons (muons or electrons) are reconstructed using the particle-flow tracks collection. Muons are required to have a transverse momentum ($p_T$) greater than 18 GeV and the pseudorapidity within the range of $|\eta| < 2.5$. Conversely, electrons are required to have $p_T$ > 22 GeV and $|\eta| < 2.5$. To ensure isolation and avoid any overlap between leptons, an overlap removal procedure is applied. This involves discarding leptons too close to each other with a distance in the $\Delta R$ less than 0.1, with the lepton having the highest $p_T$ retained.

For the $4\mu$ channel, events must have at least four muons that satisfy the previously mentioned selections. An electron veto is applied to ensure orthogonality with the $4e$ and $2e2\mu$ channels. This consists of discarding events with electrons. Additionally, events are required to have at least two same charge pairs of muon. The two pairs with the highest $p_T$ are considered candidates for the doubly-charged scalar particles.

Similarly, for the $4e$ channel, events must have a minimum of four electrons that meet the selection criteria. A muon veto is applied to ensure orthogonality with the $4\mu$ and $2e2\mu$ channels. Events must have at least two pairs of electrons with the same charge. The two pairs with the highest $p_T$ are identified as doubly-charged scalars.

In the $2e2\mu$ channel, events must contain at least two electrons and two muons. No veto is applied in this channel. Events must have at least one pair of electrons and one pair of muons with the same charge. Doubly-charged scalars are reconstructed using the pairs with the highest $p_T$.

\begin{figure*}[t!]
    \centering
    \subfloat[$4\mu$ channel][$4\mu$ channel]{\includegraphics[width=0.35\textwidth]{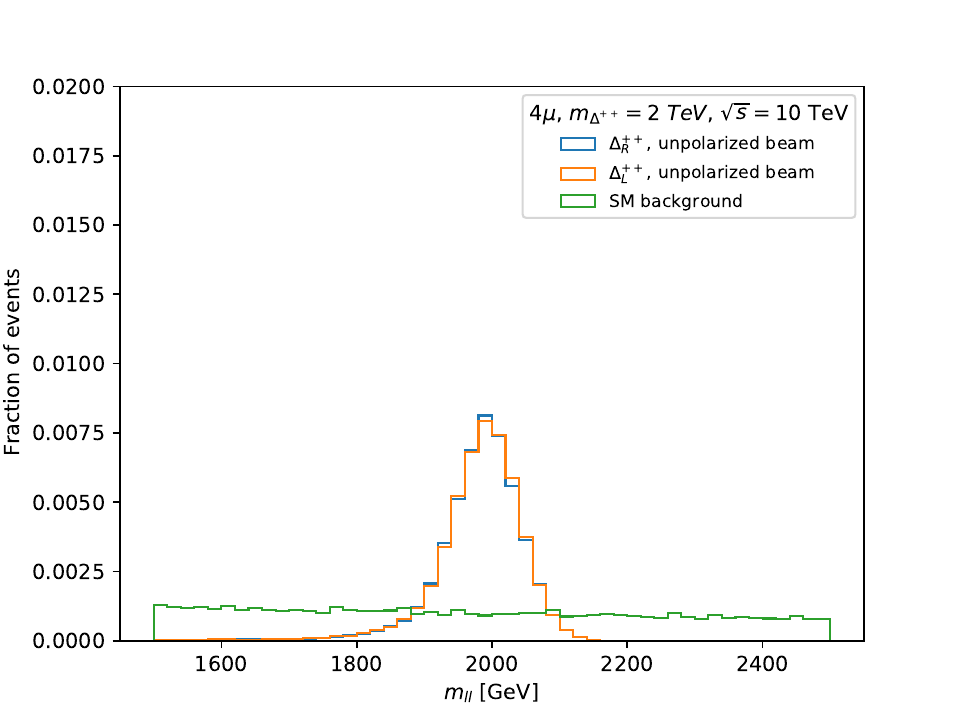}}
    \subfloat[$4e$ channel][$4e$ channel]{\includegraphics[width=0.35\textwidth]{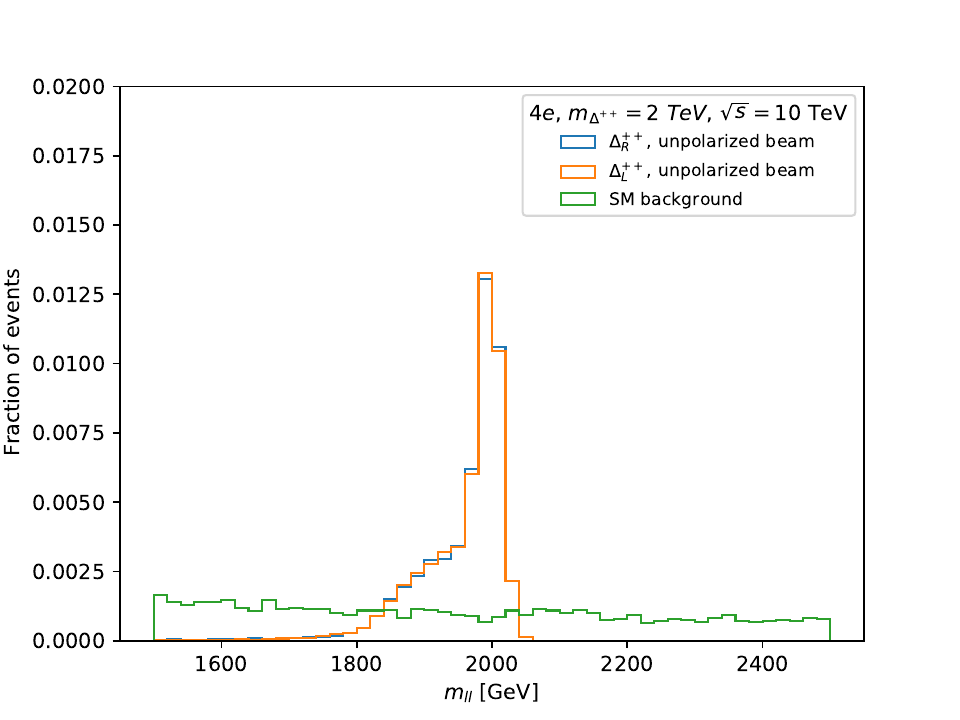}}
    \subfloat[$2e2\mu$ channel][$2e2\mu$ channel]{\includegraphics[width=0.35\textwidth]{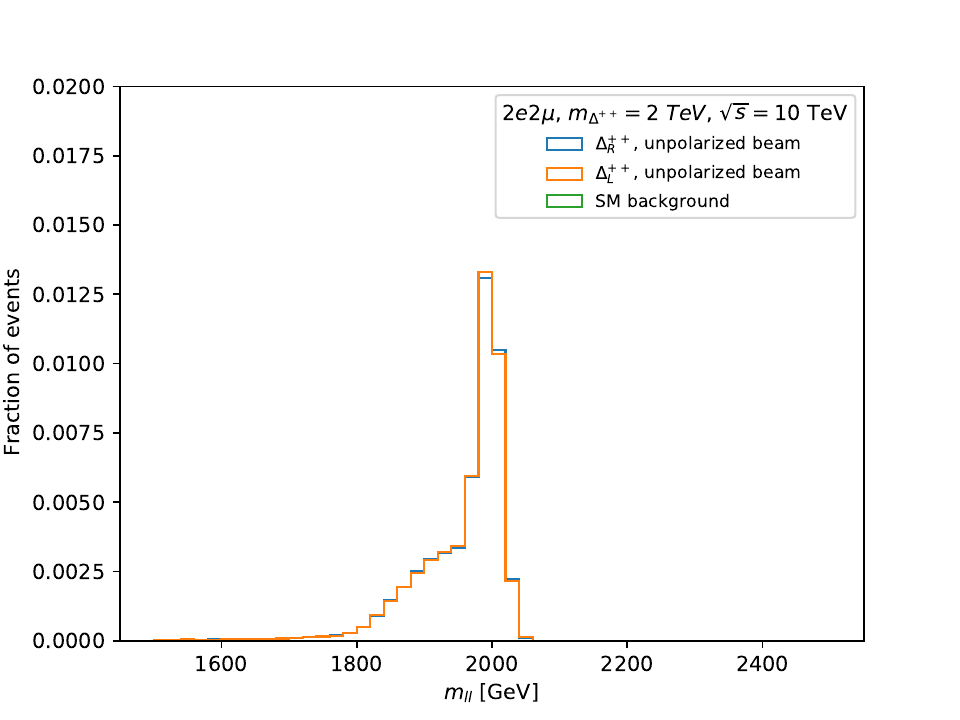}}
    \caption{The invariant mass distribution $m_{\ell\ell}$ of the reconstructed two leptons for (a) $4\mu$ channel, (b) $4e$ channel, and (c) $2e2\mu$ channel. All the Yukawa couplings are set to be 0.01.}
    \label{fig:mll}
\end{figure*}

For each selected event, the invariant masses $m_{\ell\ell}$ of each of the doubly-charged scalar candidates are computed, and events are required to satisfy that $m_{\ell\ell}$ within $ m_{\Delta^{++}} \pm 10\%$ (where $ m_{\Delta^{++}} \pm 10\%$ is the generated hypothesis). In addition, this mass requirement is removed to allow for more background events in $4e$ and $2e2\mu$ channels. Fig.~\ref{fig:mll} shows the distribution of the $m_{\ell\ell}$ invariant mas for each channel for $m_{\Delta^{++}} = $ 2 TeV. The low tails observed in the reconstructed $m_{ee}$ for both $2e2\mu$ and $4e$ channels are due to the FCC-$hh$ tracking resolution, which is part of Delphes's muon collider detector card. For the $2e2\mu$ channel, only a few background events (at low $m_{\ell\ell}$) passed our selection. The background in this channel is due to the misidentification rate estimated to be 0.1\% at the muon collider \cite{Forslund:2022xjq}. Thus, the $2e2\mu$ channel is considered nearly background-free. 

\section{Result and discussion}\label{resultsec}
We computed the acceptance times efficiency, which is defined as the number of selected events divided by the expected number of events, for each channel. The $4\mu$ channel exhibits a higher efficiency of approximately $77\%$ compared to the $4e$ and $2e2\mu$ channels. Consequently, the search for doubly-charged scalar sensitivity is predominantly driven by the $4\mu$ channel. Notably, the acceptance times efficiency remains flat across the scalar mass for both LH and RH scalars. Besides, the initial beam polarization does not influence the final state kinematics, ensuring the efficiency remains unaffected. However, as demonstrated in Figure \ref{fig:cross_vs_pol}, the rate is enhanced by a factor of 3 in for LH scalar production when a fully polarised beam is considered.

For $m_{\Delta^{++}_{L,R}} $ = 2 TeV, tables \ref{tab:dl2pPol0} and \ref{tab:dr2pPol0} are provided to summarize the number of selected signal $N_{signal}$ and SM background events $N_{SM}$ for each channel. The $2e2\mu$ channel is reported to have a very low background contribution, as described earlier. Additionally, the expected Asimov significance~\cite{Cowan:2010js} (Z) is also reported.

\begin{table*}[ht]
\centering 
\resizebox{0.95\textwidth}{!}{\begin{tabular}{|l|l|l|l|l|l|l|l|l|l|l|} 
\hline
\multirow{2}{*}{} &
  \multicolumn{5}{c|}{Unpolarized} &
  \multicolumn{5}{c|}{Polarized} \\
  \hline
Channel & $\sigma$ [fb] & $N_{signal}$ & $Acc\times\epsilon$ [\%] & $N_{SM}$ & Z & $\sigma$ [fb] & $N_{signal}$ & $Acc\times\epsilon$ [\%] & $N_{SM}$ & Z \\ 
\hline 
$4\mu$ & 0.010 & 747.219 $\pm$ 3.82 & 76.45 & 3.041 $\pm$ 0.03 & 80.235 $\pm$ 0.21 & 0.031 & 2454.285 $\pm$ 12.54 & 76.62 & 5.881 $\pm$ 0.11 & 145.431 $\pm$ 0.26\\ 
 $4e$ & 0.010 & 473.731 $\pm$ 3.04 & 48.47 & 1.087 $\pm$ 0.00 & 62.225 $\pm$ 0.09 &  0.031 & 1543.145 $\pm$ 9.94 & 48.17 & 2.350 $\pm$ 0.00 & 112.265 $\pm$ 0.11\\ 
 $2e2\mu$ & 0.019 & 1164.477 $\pm$ 6.75 & 59.57 & (78.032 $\pm$ 0.761) $\times 10^{-3}$ & 98.744 $\pm$ 0.004 & 0.062 & 3811.128 $\pm$ 22.10 & 59.49 & (150.880 $\pm$ 09.382) $\times 10^{-3}$ & 178.622 $\pm$ 0.01\\ 
 \hline 
\end{tabular}}
\caption{Summary of selected events and expected Asimov significance (Z) for $\Delta_{L}^{++}$ with mass 2 TeV at $\sqrt{s} = 10  $ TeV and $\mathcal{L} = 10\,\mathrm{ab}^{-1}$ for unpolarized and polarized beams. All the Yukawa couplings are set to be 0.01.}
\label{tab:dl2pPol0}
\end{table*}

\begin{table*}[ht]
\centering 
\resizebox{0.95\textwidth}{!}{\begin{tabular}{|l|l|l|l|l|l|l|l|l|l|l|} 
\hline
\multirow{2}{*}{} &
  \multicolumn{5}{c|}{Unpolarized} &
  \multicolumn{5}{c|}{Polarized} \\
  \hline
Channel & $\sigma$ [fb] & $N_{signal}$ & $Acc\times\epsilon$ [\%] & $N_{SM}$ & Z & $\sigma$ [fb] & $N_{signal}$ & $Acc\times\epsilon$ [\%] & $N_{SM}$ & Z \\ 
\hline  
$4\mu$ & 0.009 & 668.701 $\pm$ 3.43 & 76.19 & 3.041 $\pm$ 0.03 & 75.888 $\pm$ 0.22 & 0.007 & 538.365 $\pm$ 2.75 & 76.67 & 5.881 $\pm$ 0.11 & 68.116 $\pm$ 0.55\\ 
 $4e$ & 0.009 & 421.530 $\pm$ 2.72 & 48.03 & 1.087 $\pm$ 0.00 & 58.665 $\pm$ 0.09 & 0.007 & 339.310 $\pm$ 2.18 & 48.32 & 2.350 $\pm$ 0.00 & 52.652 $\pm$ 0.23\\ 
 $2e2\mu$ & 0.017 & 1042.117 $\pm$ 6.05 & 59.37 & (78.032 $\pm$ 0.761) $\times 10^{-3}$ & 93.393 $\pm$ 0.005 & 0.014 & 835.404 $\pm$ 4.84 & 59.49 & (150.880 $\pm$ 09.382) $\times 10^{-3}$ & 83.629 $\pm$ 0.02 \\ 
 \hline
\end{tabular}
}
\caption{ Summary of selected events and expected Asimov significance (Z) for $\Delta_{R}^{++}$ with mass 2 TeV at $\sqrt{s} = 10  $ TeV and $\mathcal{L} = 10\,\mathrm{ab}^{-1}$ for unpolarized and polarized beams. All the Yukawa couplings are set to be 0.01.}
\label{tab:dr2pPol0}
\end{table*}

When considering an unpolarized initial beam, the Asimov significance, for $\Delta^{++}_{L}\,(\Delta^{++}_{R})$, are  98.7 (93.3), 62.2 (58.6), and 80.23 (75.9) for the $2e2\mu$, $4e$, and $4\mu$ channels, respectively. The $\Delta^{++}_{L}$ exhibits a 6\% higher significance compared to the $\Delta^{++}_{R}$, primarily due to a slightly higher cross-section, as shown in Figure \ref{fig:cross_vs_pol} (a). The $2e2\mu$ shows the highest significance due to being a nearly background-free final state. However, this $6\%$ difference is significantly improved for a fully polarized beam, resulting in a remarkable factor two enhancement in the sensitivity. Thus using a polarized beam will significantly enhance the muon collider's sensitivity to probe the doubly-charged scalars of MLRSM.

\begin{figure}[t!]
    \centerline{
    \subfloat[$\Delta^{++}_{L}$][$\Delta^{++}_{L}$]{\includegraphics[width=0.28\textwidth]{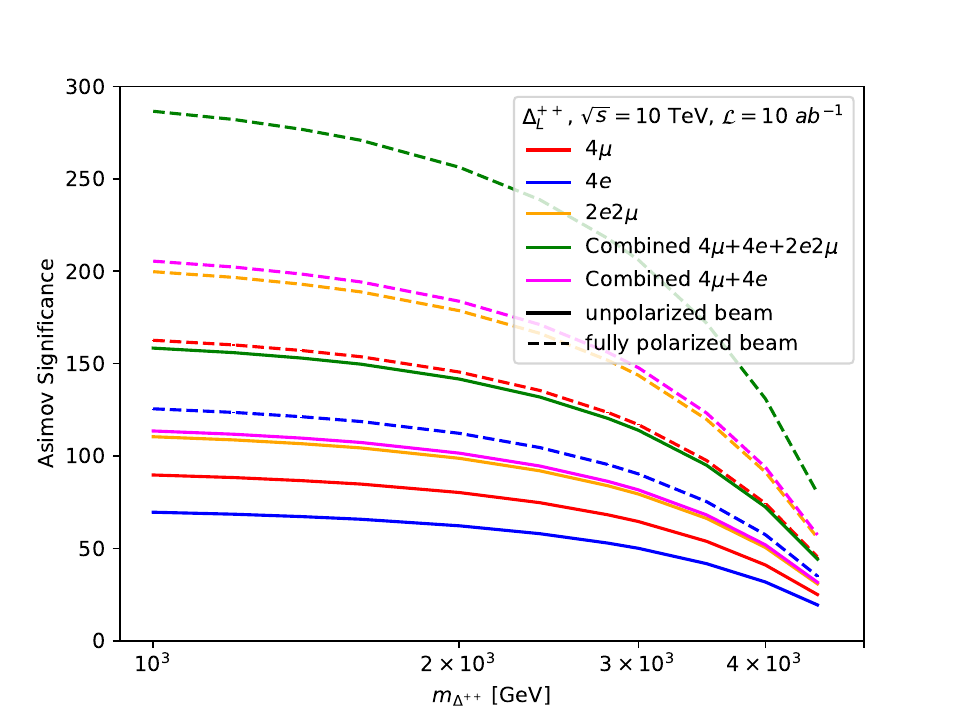}}\hspace{-1mm}
    \subfloat[$\Delta^{++}_{R}$][$\Delta^{++}_{R}$]{\includegraphics[width=0.28\textwidth]{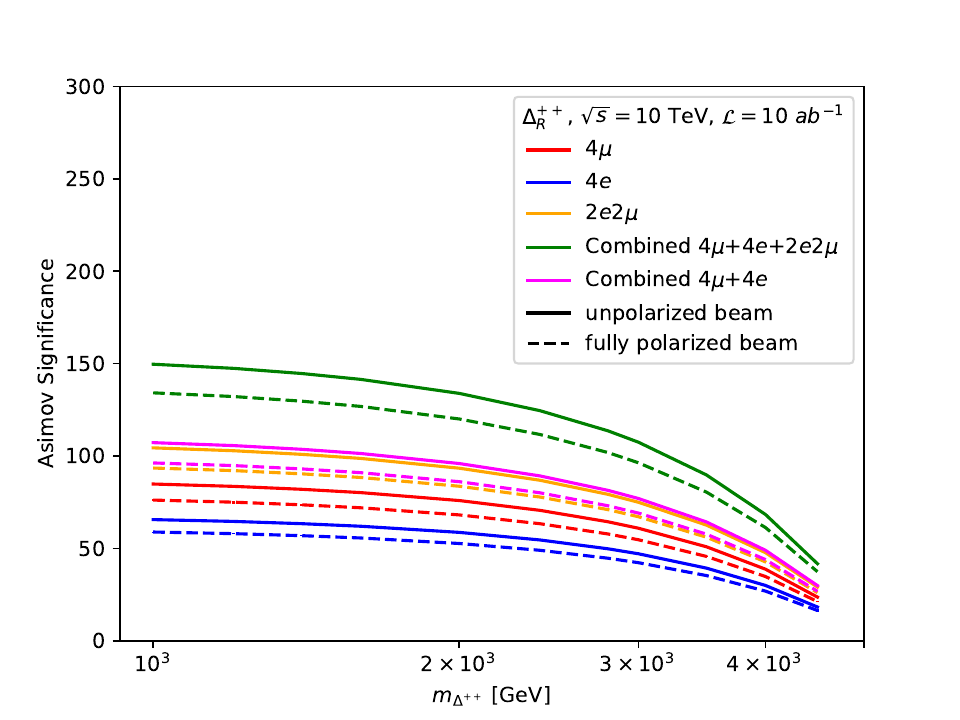}}}   
    \caption{Asimov significance of (a) $\Delta^{++}_{L}$ and (b) $\Delta^{++}_{R}$ as a function of $m_{\Delta^{++}}$, for $4\mu$ channel (red), $4e$ channel (blue) and $2e2\mu$ channel (orange). The combinations of the $4\mu$, $4e$, and $2e2\mu$ channels are presented with green lines. The combinations of $4\mu$ and $4e$ channels are presented with magenta lines. Solid (dashed) lines consist of un- (fully) polarized initial beams. All the Yukawa couplings are set to be 0.01.}
    \label{fig:sigma_summary}
\end{figure}

Figure \ref{fig:sigma_summary} illustrates the Asimov significance for the $4\mu$, $4e$ and $2e2\mu$ channels as a function of $m_{\Delta^{++}_{L,R}}$ for $\sqrt{s}=10$ TeV. As we approach the $\sqrt{s}/2$, the sensitivity decreases significantly, as expected. However, the significance remains prominent even at this point, showcasing the robustness of this very clean final state. Besides, the combined results of $4\mu$ + $4e$ + $2e2\mu$ are presented where each channel is statistically combined by ensuring the full orthogonality of different signal regions as described in the previous section. The combination is performed by summing the expected significance of each category in quadrature. This combination significantly enhances our sensitivity, achieving a factor of 1.3 improvements, as seen from green continuous (dashed) lines in Fig. \ref{fig:sigma_summary} for unpolarized (polarized) muon beams. However, since the $2e2\mu$ channel has the lowest background contribution, we present the combined results of $4\mu$ and $4e$ channels, represented by magenta continuous (dashed) lines for unpolarized (polarized) muon beams. Remarkably, we see that the combined result of $4e+4\mu$ channels is equivalent to the results of the $2e2\mu$ channel.

\begin{figure}[t!]
    \centerline{
    \subfloat[$\Delta^{++}_{L}$][$\Delta^{++}_{L}$]{\includegraphics[width=0.29\textwidth]{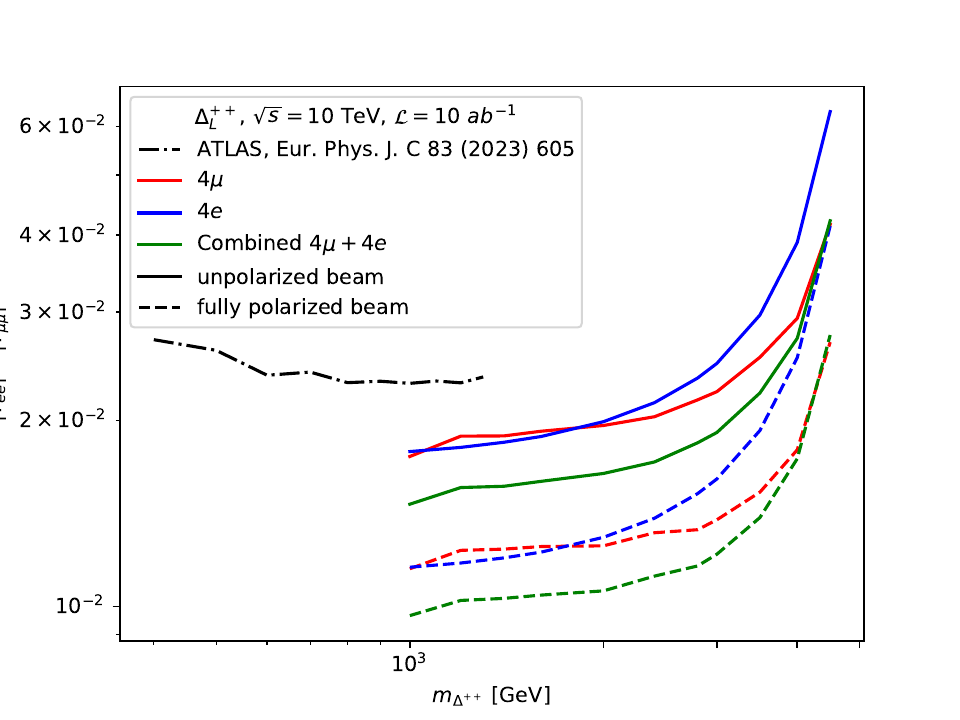}}\hspace{-1mm}
    \subfloat[$\Delta^{++}_{R}$][$\Delta^{++}_{R}$]{\includegraphics[width=0.29\textwidth]{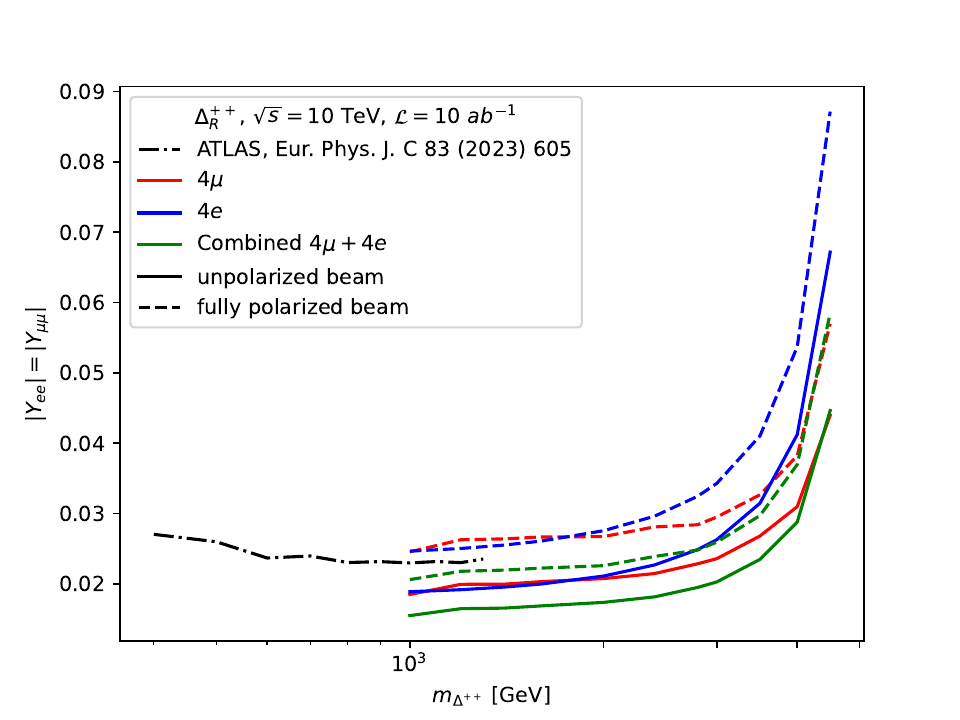}}}    
    \caption{95\% CL exclusion limit on the Yukawa couplings of (a) $\Delta^{++}_{L}$ and (b) $\Delta^{++}_{R}$ scalars as a function of their mass, for $4\mu$ channel (red), $4e$ channel (blue). The combination of $4\mu$ and $4e$ channels is presented with green lines. Solid (dashed) lines consist of an un- (fully) polarized initial beam. }
    \label{fig:limit_summary}
\end{figure}

Finally, as the cross-section, $\mu^{+}\mu^{-}\rightarrow \Delta^{++}_{L(R)}\Delta^{--}_{L(R)}\rightarrow \ell^{+}\ell^{+}\ell^{-}\ell^{-}$ depends on the Yukawa couplings, $Y_{\ell\ell}$, we computed the expected limits on their magnitudes for fixed $m_{\Delta^{++}}$ values. In Fig.~\ref{fig:limit_summary}, we present the exclusion limits at a $95\%$ confidence level for the Yukawa couplings as a function of $m_{\Delta^++}$, focusing on $4\mu$ and $4e$ channels. In the calculation of the CLs, we assume that the expected number of observed events is equal to the background expectations. Furthermore, we assume that the uncertainty on the background yield is 5\%. The $4\mu$ and $4e$ channels exhibit similar sensitivity in the low mass region. However, at higher masses around 4 TeV, the $4\mu$ channel significantly outperforms the $4e$ channel, providing a stronger constraint on the Yukawa coupling. Combining the results from the $4\mu$ and $4e$ channels, we achieve a $20\%$ improvement in our exclusion limit in the low mass region. Additionally, when examining the case of $\Delta^{++}_{L}$, considering a fully polarized beam further strengthens the constraint, unlike $\Delta^{++}_{R}$, where the sensitivity is diminished.

Besides, we include the $95\%$ CL limit on the Yukawa couplings as a function of the doubly-charged scalar mass in Fig.~\ref{fig:limit_summary} from ATLAS \cite{ATLAS:2022pbd}, derived from their exclusion limits\footnote{Available in \href{https://www.hepdata.net/record/ins2181753}{\text{https://www.hepdata.net/record/ins2181753}}} on the doubly-charged scalar production cross-sections with scalar mass. We opt to use the ATLAS result as it provides better sensitivity on the Yukawa couplings compared to the LEP searches \cite{L3:2003zst}. Moreover, the future electron-positron collider CLIC with $\sqrt{s} = 3$ TeV and $\mathcal{L}=2\,\mathrm{ab}^{-1}$ can probe doubly-charged scalar mass within 500 GeV - 2.5 TeV and Yukawa coupling within $[.001, 0.1]$ \cite{CLIC:2018fvx}, whereas from Fig.~\ref{fig:limit_summary}, we can see that the muon collider with larger $\sqrt{s}=10$ TeV and $\mathcal{L}=10\,\mathrm{ab}^{-1}$ can easily probe doubly-charged scalars beyond the reach of the CLIC.

\section{Conclusion}
\label{conclusion}
A multi-TeV muon collider can not only probe both the SM and BSM but also complement the new physics searches for future hadron colliders. Therefore, in this work, we focused on the possibility of probing the doubly-charged scalars $\Delta^{++}_{L}$ and $\Delta^{++}_{R}$ of the MLRSM in the future multi-TeV muon collider using the process, $\mu^{+}\mu^{-}\rightarrow \Delta^{++}_{L(R)}\Delta^{--}_{L(R)}\rightarrow \ell^{+}\ell^{+}\ell^{-}\ell^{-}$ with $\ell=e,\,\mu$. We choose the parameter space where masses of $W_{R}$ and flavor violating Higgses $H^{+}$, $H^{0}$ and $A^{0}$ are in $O(20)$ TeV. We concentrate on the doubly-charged mass in the range, $m_{\Delta^{++}}=1.1-5$ TeV in the muon collider with center of mass energy $\sqrt{s}=10$ TeV and integrated luminosity, $\mathcal{L}=10\,\mathrm{ab}^{-1}$. Besides, to avoid the stringent constraints coming from the lepton flavour violating processes $\mu\rightarrow 3 e$ and $\tau\rightarrow 3e,\,3\mu$, which are mediated by the exchange of doubly-charged scalars, we consider a diagonal triplet Yukawa matrix $Y$ with $Y_{ee}=Y_{\mu\mu}=Y_{\tau\tau}$. From our sensitivity analysis of probing $\Delta^{++}_{L,R}$ with four lepton final states that contain the same charge pairs in the muon collider, we point out as follows.
Due to the lowest SM background, the $2e2\mu$ channel has the largest sensitivity of probing $\Delta^{++}_{L,R}$. The next sensitive channels are the $4\mu$ and $4e$ channels. Although the production cross-sections of $\Delta^{++}_{L}$ and $\Delta^{++}_{R}$ are comparable in the case of unpolarized initial muon beams, the difference between them becomes significant when the polarized muon beams are used, and will allow us to distinguish between $\Delta^{++}_{L}$ and $\Delta^{++}_{R}$ in the muon collider. Finally, the $4\mu$ channel has a stronger exclusion limit on the Yukawa couplings at higher $m_{\Delta^{++}}$ masses. The constraints are improved with a combined analysis of $4\mu$ and $4e$ channels.

In conclusion, the doubly-charged scalars $\Delta^{++}_{L}$ and $\Delta^{++}_{R}$ are the unique predictions of the MLRSM which are connected to its explanation of neutrino mass through the interplay of left-right symmetry. The multi-TeV muon collider has the potential to discover them and provide the experimental verification of the MLRSM, thus making it a compelling avenue for future research to unravel the mysteries of neutrino mass and physics beyond the Standard Model.

\subsection*{Acknowledgement}
TAC is grateful to Kyoungchul Kong for illuminating discussions and would like to thank the High Energy Theory Group in the Department of Physics and Astronomy at the University of Kansas for their hospitality and support.  The work of S.N. is supported by the United Arab Emirates University (UAEU) under UPAR Grant No. 12S093.

\appendix
\section{Higgs potential of MLRSM}\label{mlrsm-higgs}
The Higgs potential of the parity symmetric MLRSM in the seesaw picture is given by,
\begin{align}
    &V= -\mu_{1}^{2}\mathrm{Tr}\left(\Phi^{\dagger}\Phi \right)-\mu^{2}_{2}\left[\mathrm{Tr}\left(\tilde{\Phi}^{\dagger}\Phi \right)+\mathrm{h.c.}\right]-\mu^{2}_{3}\left[\mathrm{Tr}\left(\mathbf{\Delta}_{L}\mathbf{\Delta}^{\dagger}_{L} \right)+\mathrm{Tr}\left(\mathbf{\Delta}_{R}\mathbf{\Delta}^{\dagger}_{R} \right)\right]\nonumber\\
    &+\lambda_{1}\left(\mathrm{Tr}\left(\Phi^{\dagger}\Phi \right)\right)^{2}
    +\lambda_{2}\left[\left(\mathrm{Tr}\left(\tilde{\Phi}\Phi^{\dagger} \right)\right)^{2}+\mathrm{h.c}\right]+\lambda_{3}\mathrm{Tr}\left(\tilde{\Phi}\Phi^{\dagger} \right)\,\mathrm{Tr}\left(\tilde{\Phi}^{\dagger}\Phi \right)\nonumber\\
    &+\lambda_{4}\mathrm{Tr}\left(\Phi^{\dagger}\Phi \right)\left[ \mathrm{Tr}\left(\tilde{\Phi}\Phi^{\dagger} \right)+\mathrm{h.c}\right]+
    \rho_{1}\left[\left(\mathrm{Tr}\left(\mathbf{\Delta}_{L}\mathbf{\Delta}_{L}^{\dagger} \right)\right)^{2}+\left(\mathrm{Tr}\left(\mathbf{\Delta}_{R}\mathbf{\Delta}_{R}^{\dagger} \right)\right)^{2}\right]\nonumber\\
    &+\rho_{2}\left[\mathrm{Tr}\left(\mathbf{\Delta}_{L}\mathbf{\Delta}_{L} \right)\mathrm{Tr}\left(\mathbf{\Delta}^{\dagger}_{L}\mathbf{\Delta}^{\dagger}_{L} \right)+L\rightarrow R\right]
    +\rho_{3}\mathrm{Tr}\left(\mathbf{\Delta}_{L}\mathbf{\Delta}_{L}^{\dagger} \right)\mathrm{Tr}\left(\mathbf{\Delta}_{R}\mathbf{\Delta}_{R}^{\dagger} \right)\nonumber\\
&+\rho_{4}\left[\mathrm{Tr}\left(\mathbf{\Delta}^{\dagger}_{L}\mathbf{\Delta}_{L}^{\dagger} \right)\mathrm{Tr}\left(\mathbf{\Delta}_{R}\mathbf{\Delta}_{R} \right)+\mathrm{h.c}\right]+\alpha_{1}\mathrm{Tr}\left(\Phi^{\dagger}\Phi\right)\left[\mathrm{Tr}\left(\mathbf{\Delta}_{L}\mathbf{\Delta}_{L}^{\dagger} \right)+\mathrm{Tr}\left(\mathbf{\Delta}_{R}\mathbf{\Delta}_{R}^{\dagger} \right) \right]\nonumber\\
    &+\alpha_{2}\left[ e^{i c}\left\{\mathrm{Tr}\left(\tilde{\Phi}\Phi^{\dagger} \right)\mathrm{Tr}\left(\mathbf{\Delta}_{L}\mathbf{\Delta}_{L}^{\dagger} \right)+\mathrm{Tr}\left(\Phi\tilde{\Phi}^{\dagger} \right)\mathrm{Tr}\left(\mathbf{\Delta}_{R}\mathbf{\Delta}_{R}^{\dagger} \right) \right\}+\mathrm{h.c}\right]\nonumber\\
&+\alpha_{3}\left[\mathrm{Tr}\left(\Phi\Phi^{\dagger}\mathbf{\Delta}_{L}\mathbf{\Delta}_{L}^{\dagger} \right)+\mathrm{Tr}\left(\Phi^{\dagger}\Phi\mathbf{\Delta}_{R}\mathbf{\Delta}_{R}^{\dagger} \right) \right].
    \label{higgs}
\end{align}
The masses of the component fields of $\mathrm{\Delta}_{L}$, after mass diagonalization in the limit $a,\,c\rightarrow 0$, are given as,
\begin{align}
    m^{2}_{\Delta^{++}_{L}}&= \frac{1}{2}(\rho_{3}-2\rho_{1})v_{R}^{2}+\frac{1}{2}\alpha_{3}\cos 2\beta\,v^{2},\label{scalarmass1}\\
    m^{2}_{\Delta^{+}_{L}}&\simeq\frac{1}{2}(\rho_{3}-2\rho_{1})v_{R}^{2}+\frac{1}{4}\alpha_{3}\cos 2\beta\,v^{2},\label{scalarmass2}\\
    m^{2}_{\Delta^{0}_{L}}&\simeq\frac{1}{2}(\rho_{3}-2\rho_{1})v_{R}^{2}.\label{scalarmass3}\\
\end{align}
On the other hand, the masses of the component fields of $\mathrm{\Delta}_{R}$ are
\begin{equation}
    m^{2}_{\Delta^{++}_{R}}=2 \rho_{2} v_{R}^2+\frac{1}{2}\alpha_{3}\cos 2\beta\,v^{2},\,\,m^{2}_{\mathrm{Re}\Delta^{0}_{R}}\simeq 2 \rho_{1} v_{R}^2\label{scalarmass5}.
\end{equation}
In addition, when $\rho_{2}\simeq \frac{1}{4}(\rho_{3}-2\rho_{1})$ we can have $m_{\Delta^{++}_{L}}\sim m_{\Delta^{++}_{R}}$.


\begin{thebibliography}{10}
\expandafter\ifx\csname url\endcsname\relax
  \def\url#1{\texttt{#1}}\fi
\expandafter\ifx\csname urlprefix\endcsname\relax\def\urlprefix{URL }\fi
\expandafter\ifx\csname href\endcsname\relax
  \def\href#1#2{#2} \def\path#1{#1}\fi

\bibitem{Mohapatra:1974gc}
R.~N. Mohapatra, J.~C. Pati, {A Natural Left-Right Symmetry}, Phys. Rev. D 11 (1975) 2558.
\newblock \href {http://dx.doi.org/10.1103/PhysRevD.11.2558} {\path{doi:10.1103/PhysRevD.11.2558}}.

\bibitem{Mohapatra:1974hk}
R.~N. Mohapatra, J.~C. Pati, {Left-Right Gauge Symmetry and an Isoconjugate Model of CP Violation}, Phys. Rev. D 11 (1975) 566--571.
\newblock \href {http://dx.doi.org/10.1103/PhysRevD.11.566} {\path{doi:10.1103/PhysRevD.11.566}}.

\bibitem{Senjanovic:1975rk}
G.~Senjanovic, R.~N. Mohapatra, {Exact Left-Right Symmetry and Spontaneous Violation of Parity}, Phys. Rev. D 12 (1975) 1502.
\newblock \href {http://dx.doi.org/10.1103/PhysRevD.12.1502} {\path{doi:10.1103/PhysRevD.12.1502}}.

\bibitem{Senjanovic:1978ev}
G.~Senjanovic, {Spontaneous Breakdown of Parity in a Class of Gauge Theories}, Nucl. Phys. B 153 (1979) 334--364.
\newblock \href {http://dx.doi.org/10.1016/0550-3213(79)90604-7} {\path{doi:10.1016/0550-3213(79)90604-7}}.

\bibitem{Mohapatra:1979ia}
R.~N. Mohapatra, G.~Senjanovic, {Neutrino Mass and Spontaneous Parity Nonconservation}, Phys. Rev. Lett. 44 (1980) 912.
\newblock \href {http://dx.doi.org/10.1103/PhysRevLett.44.912} {\path{doi:10.1103/PhysRevLett.44.912}}.

\bibitem{Mohapatra:1980yp}
R.~N. Mohapatra, G.~Senjanovic, {Neutrino Masses and Mixings in Gauge Models with Spontaneous Parity Violation}, Phys. Rev. D 23 (1981) 165.
\newblock \href {http://dx.doi.org/10.1103/PhysRevD.23.165} {\path{doi:10.1103/PhysRevD.23.165}}.

\bibitem{Minkowski:1977sc}
P.~Minkowski, {$\mu \to e\gamma$ at a Rate of One Out of $10^{9}$ Muon Decays?}, Phys. Lett. B 67 (1977) 421--428.
\newblock \href {http://dx.doi.org/10.1016/0370-2693(77)90435-X} {\path{doi:10.1016/0370-2693(77)90435-X}}.

\bibitem{Gell-Mann:1979vob}
M.~Gell-Mann, P.~Ramond, R.~Slansky, {Complex Spinors and Unified Theories}, Conf. Proc. C 790927 (1979) 315--321.
\newblock \href {http://arxiv.org/abs/1306.4669} {\path{arXiv:1306.4669}}.

\bibitem{Glashow:1979nm}
S.~L. Glashow, {The Future of Elementary Particle Physics}, NATO Sci. Ser. B 61 (1980) 687.
\newblock \href {http://dx.doi.org/10.1007/978-1-4684-7197-7_15} {\path{doi:10.1007/978-1-4684-7197-7_15}}.

\bibitem{Yanagida:1979as}
T.~Yanagida, {Horizontal gauge symmetry and masses of neutrinos}, Conf. Proc. C 7902131 (1979) 95--99.

\bibitem{Pati:1974yy}
J.~C. Pati, A.~Salam, {Lepton Number as the Fourth Color}, Phys. Rev. D 10 (1974) 275--289, [Erratum: Phys.Rev.D 11, 703--703 (1975)].
\newblock \href {http://dx.doi.org/10.1103/PhysRevD.10.275} {\path{doi:10.1103/PhysRevD.10.275}}.

\bibitem{Chang:1983fu}
D.~Chang, R.~N.~Mohapatra and M.~K.~Parida,
{Decoupling Parity and SU(2)-R Breaking Scales: A New Approach to Left-Right Symmetric Models}, Phys. Rev. Lett. \textbf{52}, 1072 (1984).
\newblock \href {https://doi.org/10.1103/PhysRevLett.52.1072} {\path{doi:10.1103/PhysRevLett.52.1072}}

\bibitem{Nemevsek:2012iq}
M.~Nemevsek, G.~Senjanovic, V.~Tello, {Connecting Dirac and Majorana Neutrino Mass Matrices in the Minimal Left-Right Symmetric Model}, Phys. Rev. Lett. 110~(15) (2013) 151802.
\newblock \href {http://arxiv.org/abs/1211.2837} {\path{arXiv:1211.2837}}, \href {http://dx.doi.org/10.1103/PhysRevLett.110.151802} {\path{doi:10.1103/PhysRevLett.110.151802}}.

\bibitem{Senjanovic:2016vxw}
G.~Senjanovi\'c, V.~Tello, {Probing Seesaw with Parity Restoration}, Phys. Rev. Lett. 119~(20) (2017) 201803.
\newblock \href {http://arxiv.org/abs/1612.05503} {\path{arXiv:1612.05503}}, \href {http://dx.doi.org/10.1103/PhysRevLett.119.201803} {\path{doi:10.1103/PhysRevLett.119.201803}}.

\bibitem{Senjanovic:2018xtu}
G.~Senjanovic, V.~Tello, {Disentangling the seesaw mechanism in the minimal left-right symmetric model}, Phys. Rev. D 100~(11) (2019) 115031.
\newblock \href {http://arxiv.org/abs/1812.03790} {\path{arXiv:1812.03790}}, \href {http://dx.doi.org/10.1103/PhysRevD.100.115031} {\path{doi:10.1103/PhysRevD.100.115031}}.

\bibitem{Senjanovic:2019moe}
G.~Senjanovic, V.~Tello, {Parity and the origin of neutrino mass}, Int. J. Mod. Phys. A 35~(09) (2020) 2050053.
\newblock \href {http://arxiv.org/abs/1912.13060} {\path{arXiv:1912.13060}}, \href {http://dx.doi.org/10.1142/S0217751X20500530} {\path{doi:10.1142/S0217751X20500530}}.

\bibitem{Senjanovic:2023czt}
G.~Senjanovi\'c, V.~Tello, {Spontaneous Parity Violation}, in: {8th Symposium on Prospects in the Physics of Discrete Symmetries}, 2023.
\newblock \href {http://arxiv.org/abs/2306.09512} {\path{arXiv:2306.09512}}.

\bibitem{Racah:1937qq}
G.~Racah, {On the symmetry of particle and antiparticle}, Nuovo Cim. 14 (1937) 322--328.
\newblock \href {http://dx.doi.org/10.1007/BF02961321} {\path{doi:10.1007/BF02961321}}.

\bibitem{Furry:1939qr}
W.~H. Furry, {On transition probabilities in double beta-disintegration}, Phys. Rev. 56 (1939) 1184--1193.
\newblock \href {http://dx.doi.org/10.1103/PhysRev.56.1184} {\path{doi:10.1103/PhysRev.56.1184}}.

\bibitem{Keung:1983uu}
W.-Y. Keung, G.~Senjanovic, {Majorana Neutrinos and the Production of the Right-handed Charged Gauge Boson}, Phys. Rev. Lett. 50 (1983) 1427.
\newblock \href {http://dx.doi.org/10.1103/PhysRevLett.50.1427} {\path{doi:10.1103/PhysRevLett.50.1427}}.

\bibitem{Tello:2010am}
V.~Tello, M.~Nemevsek, F.~Nesti, G.~Senjanovic, F.~Vissani, {Left-Right Symmetry: from LHC to Neutrinoless Double Beta Decay}, Phys. Rev. Lett. 106 (2011) 151801.
\newblock \href {http://arxiv.org/abs/1011.3522} {\path{arXiv:1011.3522}}, \href {http://dx.doi.org/10.1103/PhysRevLett.106.151801} {\path{doi:10.1103/PhysRevLett.106.151801}}.

\bibitem{Nemevsek:2011aa}
M.~Nemevsek, F.~Nesti, G.~Senjanovic, V.~Tello, {Neutrinoless Double Beta Decay: Low Left-Right Symmetry Scale?}\href {http://arxiv.org/abs/1112.3061} {\path{arXiv:1112.3061}}.

\bibitem{Chakrabortty:2012mh}
J.~Chakrabortty, H.~Z. Devi, S.~Goswami, S.~Patra, {Neutrinoless double-$\beta$ decay in TeV scale Left-Right symmetric models}, JHEP 08 (2012) 008.
\newblock \href {http://arxiv.org/abs/1204.2527} {\path{arXiv:1204.2527}}, \href {http://dx.doi.org/10.1007/JHEP08(2012)008} {\path{doi:10.1007/JHEP08(2012)008}}.

\bibitem{Zhang:2007fn}
Y.~Zhang, H.~An, X.~Ji, R.~N. Mohapatra, {Right-handed quark mixings in minimal left-right symmetric model with general CP violation}, Phys. Rev. D 76 (2007) 091301.
\newblock \href {http://arxiv.org/abs/0704.1662} {\path{arXiv:0704.1662}}, \href {http://dx.doi.org/10.1103/PhysRevD.76.091301} {\path{doi:10.1103/PhysRevD.76.091301}}.

\bibitem{Senjanovic:2015yea}
G.~Senjanovi\'c, V.~Tello, {Restoration of Parity and the Right-Handed Analog of the CKM Matrix}, Phys. Rev. D 94~(9) (2016) 095023.
\newblock \href {http://arxiv.org/abs/1502.05704} {\path{arXiv:1502.05704}}, \href {http://dx.doi.org/10.1103/PhysRevD.94.095023} {\path{doi:10.1103/PhysRevD.94.095023}}.

\bibitem{Nemevsek:2018bbt}
M.~Nemev\v{s}ek, F.~Nesti, G.~Popara, {Keung-Senjanovi\'c process at the LHC: From lepton number violation to displaced vertices to invisible decays}, Phys. Rev. D 97~(11) (2018) 115018.
\newblock \href {http://arxiv.org/abs/1801.05813} {\path{arXiv:1801.05813}}, \href {http://dx.doi.org/10.1103/PhysRevD.97.115018} {\path{doi:10.1103/PhysRevD.97.115018}}.

\bibitem{CMS:2021dzb}
A.~Tumasyan, et~al., {Search for a right-handed W boson and a heavy neutrino in proton-proton collisions at $ \sqrt{s} $ = 13 TeV}, JHEP 04 (2022) 047.
\newblock \href {http://arxiv.org/abs/2112.03949} {\path{arXiv:2112.03949}}, \href {http://dx.doi.org/10.1007/JHEP04(2022)047} {\path{doi:10.1007/JHEP04(2022)047}}.

\bibitem{ATLAS:2023cjo}
G.~Aad, et~al., {Search for heavy Majorana or Dirac neutrinos and right-handed $W$ gauge bosons in final states with charged leptons and jets in $pp$ collisions at $\sqrt{s}=13$ TeV with the ATLAS detector}
\newblock \href {http://arxiv.org/abs/2304.09553} {\path{arXiv:2304.09553}}.

\bibitem{ATLAS:2022pbd}
G.~Aad \textit{et al.} [ATLAS],
{Search for doubly charged Higgs boson production in multi-lepton final states using 139 fb$^{-1}$ of proton-proton collisions at $\sqrt{s}$ = 13 TeV with the ATLAS detector},
Eur. Phys. J. C \textbf{83}, no.7, 605 (2023).
\newblock \href{https://doi.org/10.1140/epjc/s10052-023-11578-9}{\path{doi:10.1140/epjc/s10052-023-11578-9}},
\href {http://arxiv.org/abs/2211.07505} {\path{arXiv:2211.07505}}.

\bibitem{Gunion:1998bc}
J.~F. Gunion, {Physics at a muon collider}, AIP Conf. Proc. 435~(1) (1998) 37--57.
\newblock \href {http://arxiv.org/abs/hep-ph/9802258} {\path{arXiv:hep-ph/9802258}}, \href {http://dx.doi.org/10.1063/1.56231} {\path{doi:10.1063/1.56231}}.

\bibitem{Palmer:2014nza}
R.~B. Palmer, {Muon Colliders}, Rev. Accel. Sci. Tech. 7 (2014) 137--159.
\newblock \href {http://dx.doi.org/10.1142/S1793626814300072} {\path{doi:10.1142/S1793626814300072}}.

\bibitem{AlAli:2021let}
H.~Al~Ali, et~al., {The muon Smasher\textquoteright{}s guide}, Rept. Prog. Phys. 85~(8) (2022) 084201.
\newblock \href {http://arxiv.org/abs/2103.14043} {\path{arXiv:2103.14043}}, \href {http://dx.doi.org/10.1088/1361-6633/ac6678} {\path{doi:10.1088/1361-6633/ac6678}}.

\bibitem{Aime:2022flm}
C.~Aime, et~al., {Muon Collider Physics Summary}\href {http://arxiv.org/abs/2203.07256} {\path{arXiv:2203.07256}}.

\bibitem{Deshpande:1990ip}
N.~G. Deshpande, J.~F. Gunion, B.~Kayser, F.~I. Olness, {Left-right symmetric electroweak models with triplet Higgs}, Phys. Rev. D 44 (1991) 837--858.
\newblock \href {http://dx.doi.org/10.1103/PhysRevD.44.837} {\path{doi:10.1103/PhysRevD.44.837}}.

\bibitem{Duka:1999uc}
P.~Duka, J.~Gluza, M.~Zralek, {Quantization and renormalization of the manifest left-right symmetric model of electroweak interactions}, Annals Phys. 280 (2000) 336--408.
\newblock \href {http://arxiv.org/abs/hep-ph/9910279} {\path{arXiv:hep-ph/9910279}}, \href {http://dx.doi.org/10.1006/aphy.1999.5988} {\path{doi:10.1006/aphy.1999.5988}}.

\bibitem{Barenboim:2001vu}
G.~Barenboim, M.~Gorbahn, U.~Nierste, M.~Raidal, {Higgs Sector of the Minimal Left-Right Symmetric Model}, Phys. Rev. D 65 (2002) 095003.
\newblock \href {http://arxiv.org/abs/hep-ph/0107121} {\path{arXiv:hep-ph/0107121}}, \href {http://dx.doi.org/10.1103/PhysRevD.65.095003} {\path{doi:10.1103/PhysRevD.65.095003}}.

\bibitem{Kiers:2005gh}
K.~Kiers, M.~Assis, A.~A. Petrov, {Higgs sector of the left-right model with explicit CP violation}, Phys. Rev. D 71 (2005) 115015.
\newblock \href {http://arxiv.org/abs/hep-ph/0503115} {\path{arXiv:hep-ph/0503115}}, \href {http://dx.doi.org/10.1103/PhysRevD.71.115015} {\path{doi:10.1103/PhysRevD.71.115015}}.

\bibitem{Tello:2012qda}
V.~Tello, {Connections between the high and low energy violation of Lepton and Flavor numbers in the minimal left-right symmetric model}, Ph.D. thesis, SISSA, Trieste (2012).

\bibitem{Dev:2016dja}
P.~S.~B. Dev, R.~N. Mohapatra, Y.~Zhang, {Probing the Higgs Sector of the Minimal Left-Right Symmetric Model at Future Hadron Colliders}, JHEP 05 (2016) 174.
\newblock \href {http://arxiv.org/abs/1602.05947} {\path{arXiv:1602.05947}}, \href {http://dx.doi.org/10.1007/JHEP05(2016)174} {\path{doi:10.1007/JHEP05(2016)174}}.

\bibitem{Maiezza:2016bzp}
A.~Maiezza, M.~Nemev\v{s}ek, F.~Nesti, {Perturbativity and mass scales in the minimal left-right symmetric model}, Phys. Rev. D 94~(3) (2016) 035008.
\newblock \href {http://arxiv.org/abs/1603.00360} {\path{arXiv:1603.00360}}, \href {http://dx.doi.org/10.1103/PhysRevD.94.035008} {\path{doi:10.1103/PhysRevD.94.035008}}.

\bibitem{Maiezza:2016ybz}
A.~Maiezza, G.~Senjanovi\'c, J.~C. Vasquez, {Higgs sector of the minimal left-right symmetric theory}, Phys. Rev. D 95~(9) (2017) 095004.
\newblock \href {http://arxiv.org/abs/1612.09146} {\path{arXiv:1612.09146}}, \href {http://dx.doi.org/10.1103/PhysRevD.95.095004} {\path{doi:10.1103/PhysRevD.95.095004}}.

\bibitem{BhupalDev:2018xya}
P.~S. Bhupal~Dev, R.~N. Mohapatra, W.~Rodejohann, X.-J. Xu, {Vacuum structure of the left-right symmetric model}, JHEP 02 (2019) 154.
\newblock \href {http://arxiv.org/abs/1811.06869} {\path{arXiv:1811.06869}}, \href {http://dx.doi.org/10.1007/JHEP02(2019)154} {\path{doi:10.1007/JHEP02(2019)154}}.

\bibitem{Beall:1981ze}
G.~Beall, M.~Bander, A.~Soni, {Constraint on the Mass Scale of a Left-Right Symmetric Electroweak Theory from the K(L) K(S) Mass Difference}, Phys. Rev. Lett. 48 (1982) 848.
\newblock \href {http://dx.doi.org/10.1103/PhysRevLett.48.848} {\path{doi:10.1103/PhysRevLett.48.848}}.

\bibitem{Mohapatra:1983ae}
R.~N. Mohapatra, G.~Senjanovic, M.~D. Tran, {Strangeness Changing Processes and the Limit on the Right-handed Gauge Boson Mass}, Phys. Rev. D 28 (1983) 546.
\newblock \href {http://dx.doi.org/10.1103/PhysRevD.28.546} {\path{doi:10.1103/PhysRevD.28.546}}.

\bibitem{Ecker:1983uh}
G.~Ecker, W.~Grimus, H.~Neufeld, {Higgs Induced Flavor Changing Neutral Interactions in SU(2)-l X SU(2)-r X U(1)}, Phys. Lett. B 127 (1983) 365, [Erratum: Phys.Lett.B 132, 467 (1983)].
\newblock \href {http://dx.doi.org/10.1016/0370-2693(83)91018-3} {\path{doi:10.1016/0370-2693(83)91018-3}}.

\bibitem{Zhang:2007da}
Y.~Zhang, H.~An, X.~Ji, R.~N. Mohapatra, {General CP Violation in Minimal Left-Right Symmetric Model and Constraints on the Right-Handed Scale}, Nucl. Phys. B 802 (2008) 247--279.
\newblock \href {http://arxiv.org/abs/0712.4218} {\path{arXiv:0712.4218}}, \href {http://dx.doi.org/10.1016/j.nuclphysb.2008.05.019} {\path{doi:10.1016/j.nuclphysb.2008.05.019}}.

\bibitem{Maiezza:2010ic}
A.~Maiezza, M.~Nemevsek, F.~Nesti, G.~Senjanovic, {Left-Right Symmetry at LHC}, Phys. Rev. D 82 (2010) 055022.
\newblock \href {http://arxiv.org/abs/1005.5160} {\path{arXiv:1005.5160}}, \href {http://dx.doi.org/10.1103/PhysRevD.82.055022} {\path{doi:10.1103/PhysRevD.82.055022}}.

\bibitem{Guadagnoli:2010sd}
D.~Guadagnoli, R.~N. Mohapatra, {TeV Scale Left Right Symmetry and Flavor Changing Neutral Higgs Effects}, Phys. Lett. B 694 (2011) 386--392.
\newblock \href {http://arxiv.org/abs/1008.1074} {\path{arXiv:1008.1074}}, \href {http://dx.doi.org/10.1016/j.physletb.2010.10.027} {\path{doi:10.1016/j.physletb.2010.10.027}}.

\bibitem{Bertolini:2014sua}
S.~Bertolini, A.~Maiezza, F.~Nesti, {Present and Future K and B Meson Mixing Constraints on TeV Scale Left-Right Symmetry}, Phys. Rev. D 89~(9) (2014) 095028.
\newblock \href {http://arxiv.org/abs/1403.7112} {\path{arXiv:1403.7112}}, \href {http://dx.doi.org/10.1103/PhysRevD.89.095028} {\path{doi:10.1103/PhysRevD.89.095028}}.

\bibitem{Bertolini:2019out}
S.~Bertolini, A.~Maiezza, F.~Nesti, {Kaon CP violation and neutron EDM in the minimal left-right symmetric model}, Phys. Rev. D 101~(3) (2020) 035036.
\newblock \href {http://arxiv.org/abs/1911.09472} {\path{arXiv:1911.09472}}, \href {http://dx.doi.org/10.1103/PhysRevD.101.035036} {\path{doi:10.1103/PhysRevD.101.035036}}.

\bibitem{SINDRUM:1987nra}
U.~Bellgardt, et~al., {Search for the Decay mu+ --> e+ e+ e-}, Nucl. Phys. B 299 (1988) 1--6.
\newblock \href {http://dx.doi.org/10.1016/0550-3213(88)90462-2} {\path{doi:10.1016/0550-3213(88)90462-2}}.

\bibitem{Hayasaka:2010np}
K.~Hayasaka, et~al., {Search for Lepton Flavor Violating Tau Decays into Three Leptons with 719 Million Produced Tau+Tau- Pairs}, Phys. Lett. B 687 (2010) 139--143.
\newblock \href {http://arxiv.org/abs/1001.3221} {\path{arXiv:1001.3221}}, \href {http://dx.doi.org/10.1016/j.physletb.2010.03.037} {\path{doi:10.1016/j.physletb.2010.03.037}}.

\bibitem{Cirigliano:2004mv}
V.~Cirigliano, A.~Kurylov, M.~J. Ramsey-Musolf, P.~Vogel, {Lepton flavor violation without supersymmetry}, Phys. Rev. D 70 (2004) 075007.
\newblock \href {http://arxiv.org/abs/hep-ph/0404233} {\path{arXiv:hep-ph/0404233}}, \href {http://dx.doi.org/10.1103/PhysRevD.70.075007} {\path{doi:10.1103/PhysRevD.70.075007}}.

\bibitem{Alloul:2013bka}
A.~Alloul, N.~D. Christensen, C.~Degrande, C.~Duhr, B.~Fuks, {FeynRules 2.0 - A complete toolbox for tree-level phenomenology}, Comput. Phys. Commun. 185 (2014) 2250--2300.
\newblock \href {http://arxiv.org/abs/1310.1921} {\path{arXiv:1310.1921}}, \href {http://dx.doi.org/10.1016/j.cpc.2014.04.012} {\path{doi:10.1016/j.cpc.2014.04.012}}.

\bibitem{Alwall:2014hca}
J.~Alwall, et~al., {The automated computation of tree-level and next-to-leading order differential cross sections, and their matching to parton shower simulations}, JHEP 07 (2014) 079.
\newblock \href {http://arxiv.org/abs/1405.0301} {\path{arXiv:1405.0301}}, \href {http://dx.doi.org/10.1007/JHEP07(2014)079} {\path{doi:10.1007/JHEP07(2014)079}}.

\bibitem{Bierlich:2022pfr}
C.~Bierlich, et~al., {A comprehensive guide to the physics and usage of PYTHIA 8.3}\href {http://arxiv.org/abs/2203.11601} {\path{arXiv:2203.11601}}, \href {http://dx.doi.org/10.21468/SciPostPhysCodeb.8} {\path{doi:10.21468/SciPostPhysCodeb.8}}.

\bibitem{Yu:2017mpx}
D.~Yu, M.~Ruan, V.~Boudry, H.~Videau, {Lepton identification at particle flow oriented detector for the future $e^{+}e^{-}$ Higgs factories}, Eur. Phys. J. C 77~(9) (2017) 591.
\newblock \href {http://arxiv.org/abs/1701.07542} {\path{arXiv:1701.07542}}, \href {http://dx.doi.org/10.1140/epjc/s10052-017-5146-5} {\path{doi:10.1140/epjc/s10052-017-5146-5}}.

\bibitem{deFavereau:2013fsa}
J.~de~Favereau, C.~Delaere, P.~Demin, A.~Giammanco, V.~Lema\^\i{}tre, A.~Mertens, M.~Selvaggi, {DELPHES 3, A modular framework for fast simulation of a generic collider experiment}, JHEP 02 (2014) 057.
\newblock \href {http://arxiv.org/abs/1307.6346} {\path{arXiv:1307.6346}}, \href {http://dx.doi.org/10.1007/JHEP02(2014)057} {\path{doi:10.1007/JHEP02(2014)057}}.

\bibitem{delphs_muon_card}
M.~Selvaggi, \href{https://indico.cern.ch/event/957299/contributions/4023467/attachments/2106044/3541874/delphes_card_mucol_mdi_.pdf}{Talk at mdi studies meeting of the muon collider collaboration}.
\newline\urlprefix\url{https://indico.cern.ch/event/957299/contributions/4023467/attachments/2106044/3541874/delphes_card_mucol_mdi_.pdf}

\bibitem{leogrande2019delphes}
E.~Leogrande, P.~Roloff, U.~Schnoor, M.~Weber, A delphes card for the clic detector (2019).
\newblock \href {http://arxiv.org/abs/1909.12728} {\path{arXiv:1909.12728}}.

\bibitem{Selvaggi:2717698}
M.~Selvaggi, \href{https://cds.cern.ch/record/2717698}{{A Delphes parameterisation of the FCC-hh detector}}, Tech. rep., CERN, Geneva (2020).
\newline\urlprefix\url{https://cds.cern.ch/record/2717698}

\bibitem{Forslund:2022xjq}
M.~Forslund, P.~Meade, {High precision higgs from high energy muon colliders}, JHEP 08 (2022) 185.
\newblock \href {http://arxiv.org/abs/2203.09425} {\path{arXiv:2203.09425}}, \href {http://dx.doi.org/10.1007/JHEP08(2022)185} {\path{doi:10.1007/JHEP08(2022)185}}.

\bibitem{Cowan:2010js}
G.~Cowan, K.~Cranmer, E.~Gross, O.~Vitells, {Asymptotic formulae for likelihood-based tests of new physics}, Eur. Phys. J. C 71 (2011) 1554, [Erratum: Eur.Phys.J.C 73, 2501 (2013)].
\newblock \href {http://arxiv.org/abs/1007.1727} {\path{arXiv:1007.1727}}, \href {http://dx.doi.org/10.1140/epjc/s10052-011-1554-0} {\path{doi:10.1140/epjc/s10052-011-1554-0}}.

\bibitem{L3:2003zst}
P.~Achard \textit{et al.} [L3],
{Search for doubly charged Higgs bosons at LEP}, Phys. Lett. B \textbf{576}, 18-28 (2003),
\newblock \href{https://doi.org/10.1016/j.physletb.2003.09.082}{\path{doi:10.1016/j.physletb.2003.09.082}}, \href{https://arxiv.org/abs/hep-ex/0309076}{\path{arXiv:hep-ex/0309076 [hep-ex]}}.

\bibitem{CLIC:2018fvx}
J.~de Blas \textit{et al.} [CLIC], {The CLIC Potential for New Physics},
\newblock \href{https://doi.org/10.23731/CYRM-2018-003}{\path{doi:10.23731/CYRM-2018-003}}, \href{https://arxiv.org/abs/1812.02093}{\path{arXiv:1812.02093 [hep-ph]}}.

\end{thebibliography}
\end{document}